\documentclass[12pt]{article}
\usepackage{graphicx}
\usepackage{amssymb,amsmath,amsfonts,palatino,amsthm}
\usepackage{amssymb}
\usepackage{epstopdf}
\usepackage{slashed} 
\newcommand{\f}{\begin{equation}}
\newcommand{\ff}{\end{equation}}
\newcommand{\blankline}{\vskip .3cm}
\begin{document}

%%%%%%%%%%%%%%%%%%%%%%%%%%%%%%%%%%%%%%%%%%%%%%%%
\title{MOND as a regime of quantum gravity \\}

\author{Lee Smolin\thanks{lsmolin@perimeterinstitute.ca} 
\\
\\
Perimeter Institute for Theoretical Physics,\\
31 Caroline Street North, Waterloo, Ontario N2J 2Y5, Canada}
\date{\today}
\maketitle
%\vfill

\begin{abstract}
 
We propose that there is a regime of quantum gravity phenomena, for the case that the cosmological constant is small and positive, which concerns physics at temperatures below the deSitter temperature, or length scales larger than the horizon.  We observe that the standard form of the equivalence principle does not apply in this regime; we consider instead that a weakened form of the equivalence principle might hold in which the ratio of gravitational to inertial mass is a function of environmental parameters.  We consider possible principles to determine that function.  These lead to behaviour that, in the limit of $\hbar \rightarrow 0$ and $c \rightarrow \infty$, reproduces the modifications of Newtonian dynamics first proposed by Milgrom.  Thus $MOND$ is elucidated as coding the physics of a novel regime of quantum gravity phenomena.

We propose also an effective description of this regime in terms of a bi-metric theory, valid in the approximation where the metric is static.  
This predicts a new effect, which modifies gravity for radial motions.

\blankline
\blankline

%{\it Dedicated to the memory of Vera Rubin}

\blankline
\blankline

%"Could be brave, or just insane....we'll have to see." 

%-{\it Another day of sun}

\end{abstract}

\newpage

\tableofcontents

%\newpage

\section{Introduction}

As discovered by Vera Rubin and her colleagues\cite{Vera}, and confirmed by many subsequent observations, the rotational velocities of stars and gas in the outer regions of spiral
galaxies depart from the $\frac{1}{\sqrt{r}}$ behavior we would expect given Newtonian gravity and the observed distributions of baryonic matter.  Instead, the velocities flatten out to constant values, $v$, given by a simple function of the total baryonic mass of the galaxy,
$M_b$,
\f
v^4 = G a_0 M_b.
\ff
where $a_0$ is an acceleration scale, which can be read off the data to be
\f
a_0=1.2 \times 10^{-8} cm/s^2  .
\ff
This empirical relation is called the {\it Baryonic} Tully-Fisher law\cite{TF}.  There are several
remarkable features of this law.

\begin{itemize}

\item{}There is remarkably little scatter, given that this is a summary of astronomical data, and galaxies are messy objects, with strongly non-linear dynamics and histories\cite{small-scatter}.  The relation appears to hold in a diverse range of disk galaxy types.

\item{}The value of $a_0$ appears universal.

\item{}$a_0$ is close to the acceleration of the universe, $a_\Lambda = c^2 \sqrt{\Lambda}$,
\f
a_0 \approx \frac{a_\Lambda}{8.3}
\ff

\item{}The relation involves the {\it baryonic} mass of a galaxy.

\end{itemize}

In 1983 Milgrom proposed\cite{MOND1} that the discrepancy of rotational velocities from Newtonian expectations could be expressed by a universal relation between
the measured radial acceleration
\f
a_{obs}= \frac{v^2}{r}
\label{v2r}
\ff
and the acceleration predicted by Newtonian theory on the basis of the observed baryonic masses,
\f
a_{N}^i = \nabla^i U
\ff
of the form
\f
a_N = a_{obs}  G^{-2} (\frac{a_{obs}}{a_0})
\ff
We can invert this to find a function $F^2 (\frac{a_{N}}{a_0})$ such that
\f
\frac{a_{obs}}{a_{N}}= F^2  (\frac{a_{N}}{a_0})
\label{ratio}
\ff
For small accelerations, compared to $a_0$, Milgrom proposed that this be chosen to reproduce the Tully-Fisher relation.  This requires that for small $a_N < < a_0$
\f
a_{obs} = \sqrt{a_N a_0}
\label{sqrt}
\ff

On the other hand, for large $a$, Newtonian gravity should be recovered.

\begin{figure}[t!]
\centering
\includegraphics[width= 0.5\textwidth]{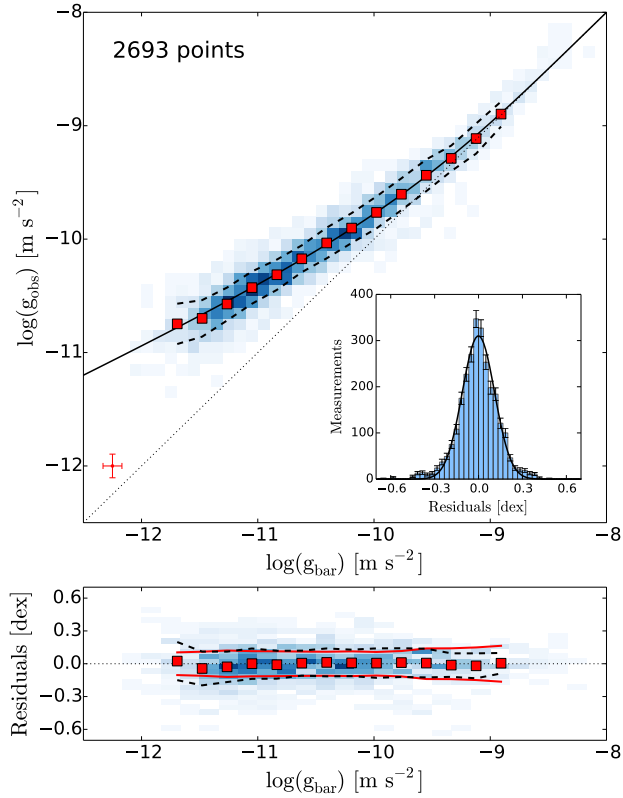}
\caption{The empirical radial acceleration relation, as shown in Figure 3 of \cite{MLS}.  Data is taken from the SPARC data base 
\cite{SPARC}. 
%{From the original caption:
%{\it ``The centripetal acceleration observed in rotation curves, $a_{obs}= g_{obs} = v^2/R$, is plotted against that predicted for the observed distribution of baryons, $a_N=g_b$ in the upper panel. Nearly 2700 individual data points for 153 SPARC galaxies are shown in grayscale. The mean uncertainty on individual points is illustrated in the lower left corner. Large squares show the mean of binned data. Dashed lines show the width of the ridge as measured by the rms in each bin. The dotted line is the line of unity. The solid line is the fit of eq. (\ref{function}),to the unbinned data using an orthogonal- distance-regression algorithm that considers errors on both variables.}  The caption goes on to say,  {\it The inset shows the histogram of all residuals and a Gaussian of width $\sigma = 0.11 dex$. The residuals are shown as a function of $g_{obs}$ in the lower panel. The error bars on the binned data are smaller than the size of the points. The solid lines show the scatter expected from observational uncertainties and galaxy to galaxy variation in the stellar mass-to-light ratio. This extrinsic scatter closely follows the observed rms scatter (dashed lines): the data are consistent with negligible intrinsic scatter.\cite{MLS}"}.} }
%Used with permission.  
}
\label{data} 
\end{figure}

In a recent paper, McGaugh, Lelli and Schombert (MLS) report\cite{MLS} strong confirmation  of an empirical relation of 
this form, first proposed by Milgrom\cite{MOND1}.
They measure
$F(\frac{a_N}{a_0})$ in a survey of rotation curves of $153$ galaxies in the SPARC data base\cite{SPARC}.  They measure 
$a_{obs}$, the actual radial acceleration by (\ref{v2r})
at $2693$ radii on these rotation curves.  At the same radii they estimate the Newtonian gravitational potential from baryons as observed in stars, 
gas and dust, and so determine $a_N $.
They discover that the data is well described by a simple empirical relation of the form of (\ref{ratio}), as shown
in Figure 1.   As they note, it is amazing that such a relation exists over a wide range of galaxy types, sizes and morphologies, as this represents the observed accelerations only by a function of the Newtonian accelerations due to baryons.  

Furthermore MLS are able to
fit a simple form for $F(a)$ (previously proposed in \cite{MS}), to the data which is\cite{MLS,Stacy08}
\f
F^2 (a_N )= \frac{1}{1-e^{-\sqrt{\frac{a_N}{a_0}}}}
\label{function}
\ff

In addition to this galactic phenomena, there is evidence for departures from Newtonian gravity on still larger scales, in observations of galactic clusters and large scale structure.  These diverse  phenomena suggests two  hypotheses.
\begin{enumerate}

\item{}{\bf Dark matter:}  Much of the matter in the universe is non-luminous and, likely, non-baryonic.

\item{}{\bf Gravity is modified}  in a regime of small acceleration. 

\end{enumerate}

Diverse theories have been introduced as elaborations on one or the other 
hypotheses.  Two remarks need to be made.  First, these are not mutually
exclusive, as the galactic and extragalactic phenomena may have different explanations.
In the early $1900$'s there were two anomalies observed in the planetary orbits; both Mercury and Neptune were off track.  The first of these was explained by a modification
of gravity, the latter by the discovery of heretofore dark matter, aka Pluto\cite{Joao-personal}.

Second, determined efforts have failed to result in a non-gravitational detection or production of dark matter.

The hypothesis that the galactic scale phenomena are explained by a modification of Newtonian gravity
was first made by Milgrom in 1983\cite{MOND1}, 
who suggested that the universal relation (\ref{ratio},\ref{sqrt},\ref{function}) he proposed was
due to the action of a fundamental modification in the laws of gravity.  

Many years later, it appears to be fair to say that at the scale of galaxies,  {\it  Milgrom's hypothesis has held up well, when compared with data\cite{heldup}.} Indeed, Milgrom's 
hypothesis must be credited with a number of predictions which were subsequently
confirmed, such as the behaviour of low surface brightness and dwarf 
galaxies\footnote{For reviews of $MOND$, see \cite{MOND-review}, some other relevant papers on MOND are \cite{Stacy}-\cite{saddle}.}.  To the extent to which
this is true, this  represents a challenge to the dark matter hypothesis.  While it is certainly conceivable that the success of Milgram's  empirical relation 
 (\ref{ratio},\ref{sqrt},\ref{function})  might be explained by a dark matter model, that model would have to explain,  
\begin{enumerate}

\item{}Why is the scatter in the Tully Fisher relation smaller when expressed in terms
of the baryonic mass than in terms of the presumed total mass\cite{small-scatter}?

\item{}Why does the acceleration relation hold widely, with small scatter, over many galaxy types, when the baryonic and dark matter distributions are weakly coupled?

\item{}Why do specific features of individual galactic rotation curves reflect the observed baryonic matter distributions, if most of the mass is dark matter, only weakly coupled to baryons?

\item{} Where does the observed acceleration scale $a_0$ come from, why is it universal, and why is it close to the cosmological acceleration?

\end{enumerate}

At the same time, as soon as one looks beyond the context of galactic rotation curves,
the $MOND$ hypothesis faces severe challenges of its own.

\begin{itemize}

\item{}It is so far expressed as a modification of non-relativistic, Newtonian dynamics.
While there are attempts to embed $MOND$ in a special or general relativistic
framework\cite{teves,moffat}, none so far are compelling theoretically.  One issue is stability, due to the 
incorporation of  a non-gauge invariant vector field.  Another issue is non-locality,
which appears necessary to code a potential that falls off slower than inverse distance.

\item{}These extended versions of $MOND$ do not do well explaining the observed
behaviours of galactic clusters, lensing, and large scale structure.

\item{}Given the acceleration relation, the theoretical proposal is underdetermined. One could regard $MOND$ either as a modification of the gravitational field equations or as a universal modification of the principle of inertia, for small acceleration.   The evidence also underdetermines the theory, because we mostly observe the low acceleration regime by steady, circular motion.
% although there are indications of the need for dark matter or modified gravity also for elliptical galaxies.

\end{itemize}

It is the aspect of non-locality, in particular, that suggests that in the, still unlikely, case that $MOND$ is true, it expresses a fundamental departure from known principles.  This of course seems unlikely, but we may note several interesting aspects of this problem.  

First, we have
no other probe of the regime of ultra-small accelerations, $a < a_\Lambda$, except the motions of stars and gas in galaxies.  

Second, we may note that an accelerating observer has, in Minkowski spacetime, an acceleration horizon at a distance,
$l_a = \frac{c^2}{a}$.  This is among other things, the peak wavelength of the corresponding Unruh radiation\cite{Unruh}.  A weakly, but uniformly,  accelerating detector is then, possibly, a very large object, at least from the viewpoint of its effect on the vacuum of the quantum fields.  It is then very interesting that the regime $a < a_\Lambda$ is also
the regime when the peak Unruh wavelength $l_a$ is of the order of the distance to the cosmological horizon, $R$, defined by $\frac{\Lambda}{3}= \frac{1}{R^2}$.

Recent research has highlighted the far infrared physics 
of gauge and gravitational physics\cite{soft}. We suggest that $MOND$ may be a surprise hiding in the non-linear dynamics of these soft modes on the scale of the cosmological horizon.

Third, we expect that fundamental phenomena, related to quantum gravity, are to be expressed in terms of 
all four of the fundamental constants, $\hbar, G, c$ and $\Lambda$.  But it is conceivable that for
$a_\Lambda$ fixed there is a regime of phenomena delineated  by $a_\Lambda$ which survives a limit in which $\hbar \rightarrow 0$ and $c \rightarrow \infty$. (Indeed the peak wavelength of Unruh radiation is a quantum phenomena in which  two $\hbar$'s have cancelled.) We may call
this, very tentatively, the cosmological constant dominated regime of quantum gravity,
for it concerns length and acceleration scales comparable to $R$ and $a_\Lambda$.

We proceed to investigate what can be said about such a regime of quantum gravity\footnote{The idea that $MOND$ is an expression of quantum gravity has been considered earlier from various points of view by Milgrom\cite{MM-related}, van Putten\cite{vanP}, Verlinde\cite{EVdark}, Hossenfelder\cite{SHonEV}, Woodard\cite{Woodard}, 
Modesto and Randano\cite{AL}, Minic et al\cite{minic},
Hendi and Sheykhi\cite{Hendi}, Mike McCulloch\cite{MM} 
and others\cite{MOND-QGothers}, including the author\cite{4principles,Leeun}.  The argument below is, I believe, on the whole, novel, but in places it overlaps some of these discussions.}.

\section{The positive cosmological constant dominated regime of quantum gravity}

What ever is the right quantum theory of quantum gravity, it will depend on four
dimensional constants, $\hbar, G, c$ and $\Lambda$.  Different regimes of quantum gravity
phenomena can be captured and delineated by studying limits such as
$\hbar \rightarrow 0$ that must recover general relativity and $G, \Lambda \rightarrow 0$
that recovers\footnote{from a perturbative perspective.} quantum field theory on flat spacetime.  The holographic regime explored by the $AdS/CFT$ correspondence is for negative $\Lambda$ with 
$\frac{1}{N}= \hbar G |\Lambda |$ small\cite{AdSCFT}.

In the last few years, new regimes have been explored, such as the relativity
locality regime in which $G$ and $\hbar$ are both taken to zero, but holding the Planck
energy, given by their ratio, fixed\cite{RL1,RL2}.

We want to explore the physics of a novel regime of physical phenomena associated
with length scales, or wavelengths greater than or comparable to $R$, and accelerations
small compared to $a_\Lambda$.  This new regime involves novel physical phenomenon which survive the limits $\hbar \rightarrow 0$ and $c \rightarrow \infty.$

What principles might govern such phenomena?

One that does not directly is the equivalence principle.  To see why,  let us review it.

As presented by Einstein, the equivalence principle consists of two statements;

{\bf EP1:}  Freely falling observers, whose extent, $L$, is much smaller than radius of curvature, $\cal R$, (and hence $R$)  observe special relativity to 
hold to zeroth order in $\frac{L}{\cal R}$. i.e. the zeroth order effects of gravity can be eliminated by free fall.

{\bf EP2:}  Uniformly accelerated observers see themselves to be in a uniform gravitational field, to zeroth order in $\frac{L}{\cal R}$.  i.e. gravity can be mocked up by uniform acceleration, to zeroth order in $\frac{L}{\cal R}$.
 
 These are equivalent classically, so long as we respect the restriction to phenomena on scales much smaller than the radius of curvature.
 
 The equivalence principle has something to say about the relationship of $m_I$-the inertial mass, to the gravitational mass, $m_g$.  The first may be defined by invoking the conservation of momentum in scattering experiments. 
 The inertial mass is then defined (for $v<<c$) as the ratio of momentum to velocity. The gravitational mass is defined independently, as a measure of the strength by which a body is affected by the gravitational field.
 
 The universality of free fall is normally taken to imply $m_g = m_I$.  But consider the possibility that free fall is universal at a given time and place, but that the response of a body to the gravitational field could differ over time and space, as a function of environmental or other parameters.   In this case it is natural to extend the principle of equivalence from $m_g=m_I$ to\cite{4principles}
\f
 \frac{m_I}{m_g} = Z(\mbox{universal function of environment, independent of masses}) .
\ff
This suffices to let us transform gravity away by going into free fall, or mock gravity up by uniformly accelerating.  This defines a universal function $Z$ which can depend on global
or environmental parameters, but does not depend on the masses themselves.

To deduce more, we have to add that the physics seen by a freely falling observer includes
Newtonian gravity in the limit of small velocities.   This then implies that
\f
m_g = m_I, \ \ \ \ \mbox{so} \ \ Z=1.
\ff

Now notice that while, consistent classically, there is a tension between the two parts of the equivalence principle.  This has to do with the different ways in which one  can try to extend them from a limiting case for which 
$\frac{L}{\cal R} \rightarrow 0$.     

The tension is due to the fact that EP1 becomes exact in the limit $L << {\cal R}$, whereas 
EP2 becomes exact in the limit of perfectly uniform acceleration, i.e. it requires not a small observer, but a static spacetime.  But uniform acceleration  implies a length
$l_a = \frac{c^2}{a}$.   This is the distance from the detector to the horizon created by the detector's uniform acceleration.
%$l_a > > L$.  

This tension becomes a conflict  quantum mechanically, because that length is the peak wavelength of the Unruh radiation, which is then the scale over which the accelerating detector
disturbs the vacuum.

We already know that the equivalence principle does not easily co-exist with quantum field theory because the  vacuum state of a quantum field is not a local object.  One cannot localize a quanta of a massive field to smaller than its Compton wavelength, $\lambda_C = \frac{\hbar}{mc}$.  But the Unruh effect makes this conflict more fraught.  

For a massless field, the Unruh effect involves for small acceleration an arbitrarily low
temperature, 
\f
T_U= \frac{\hbar a}{2 \pi c}
\ff
whose corresponding thermal distribution has a peak wavelength,
\f
\lambda_U = \frac{\hbar c}{T_U} = l_a = \frac{c^2}{a}
\ff
Thus, the limit of small acceleration or low temperature involves arbitrarily long length scales,
which brings us into conflict with the condition required by the equivalence principle
that the phenomena associated with the detector has to be smaller than the radius of curvature.

%But what about small acceleration?  Small means such that
%$l_a > R$ or $\cal R$.  But this contradicts the condition that the disturbances caused by the detector are much smaller than the radius of curvature.  

EP1 requires that all lengths are much smaller than $\cal R$ and hence are smaller than
$R$.  We are consistent with this only if the acceleration is large enough that the peak wave length of the Unruh radiation is much smaller than the radius of curvature, and hence $R$. 
This defines a phase of quantum gravity, dominated by EP1.

But  there could be another phase of quantum gravity not dominated by EP1 for small acceleration, $l_a > R$ or $a < a_\Lambda$.  Here, a restricted form of EP2 can hold, but not EP1.

Thus, when there is a positive cosmological constant, the resulting length scale, $R$, and acceleration scale, $a_\Lambda$ divide physical phenomena into two regimes.

\begin{itemize}

\item{}{\bf The equivalence principle dominated regime.}  Those phenomena, all of whose length  are less than $R$, and all of whose accelerations are greater than $a_\Lambda$, fall into  the normal regime dominated by the equivalence principle.   The classical description of this physics is general relativity.

\item{}{\bf The cosmological constant dominated regime}  includes phenomena whose length scales are greater than $R$ and/or whose acceleration scales are less than $a_\Lambda$.

\end{itemize}

One way to delineate the two regimes is to express it in terms of  the relationship between the inertial and gravitational masses.
\f
\frac{m_I}{m_g}= Z
\ff
In the equivalence principle dominated regime, $Z=1$.  In the $\Lambda$ dominated regime this can depend on other, global or environmental quantities (but not the masses themselves.)

We need a principle to determine how $Z$ depends on various quantities.   The idea is that all the non-local and  far infrared physics connected with the cosmological horizon scale are reflected in the dependence of $Z$ on global or environmental parameters.  Thus, so far as local physics is concerned, we may apply classical reasoning to the physics in the cosmological constant dominated regime, with the exception of the renormalization of the ratio between the gravitational and inertial masses of a body.

\section{The thermal equivalence principle}

What should an observer in the cosmological constant dominated regime observe?  Guidance is provided by the crucial result of  Narnhofer and Thirring\cite{NT} and Deser and Levine\cite{DL}, who find that a uniformly accelerating detector in deSitter spacetime observes a thermal spectrum with a temperature,
\f
T_{DL}= \sqrt{T_{dS}^2 + \left ( \frac{\hbar a }{2 \pi c} \right )^2 }
\label{DL}
\ff
where the deSitter temperature is
\f
T_{dS} = \frac{\hbar c^2}{2 \pi}\sqrt{\frac{\Lambda}{3}}
\ff

Note that $T_{DL} \geq T_{dS}$, so the latter is a minimum temperature for  equilibrium.

The peak wavelength is then
\f
\lambda_{DL}= 
\frac{1}{\sqrt{\frac{1}{l_{a}^2 }+    \frac{1}{R^2} }}
\ff

The peak wavelength is a pure quantum phenomenon, but so is the Unruh temperature, which means that both 
are proportional to $\hbar$.  So, when we compute the peak wavelength, the $\hbar$'s cancel and we are left with a seemingly classical
criteria marking a boundary of a phase of a quantum phenomenon:
\f
a < a_\Lambda
\ff
For $a >> a_\Lambda$,  $\lambda_U \approx l_{acc} << R$ and we are in the equivalence principle dominated phase.  But for  $a < a_\Lambda$,  $\lambda_U \approx R $ and we are in the new,
cosmological constant dominated phase.   Indeed not only do the $\hbar$'s cancel, but, when expressed in terms of accelerations, the $c$'s also cancel, so we have the possibility of a quantum gravity effect modifying Newtonian dynamics.  

Now we  don't live in an exact deSitter spacetime, so how is the Deser-Levin temperature relevant?  We propose that it apply to an observer in the cosmological constant dominated regime, for $\Lambda >0$.  

To be precise,
%{\bf The  thermal equivalence principle  (TEP)}, is restricted
let us restrict ourselves 
to static observers, whose worldlines are generated by timelike killing fields of static spacetimes.
%We assume the observer is small compared to the local radius of curvature. 
Such an observer observes an acceleration, $a^a$ and a temperature, $T$. The $T$ is a minimal temperature they can detect, when all sources of thermal radiation, including black hole horizons, but not counting the deSitter temperature, are turned off or shielded.     $a^a$ is the acceleration they observe between their trajectory and the trajectory of a freely falling body at their location.  

The {\bf thermal equivalence principle (TEP)} asserts that {\it the temperature $T$, experienced by the static detector is related to the magnitude of the acceleration, $a$, by the Deser-Levin formula
(\ref{DL}).}

This incorporates the idea of the universality of free fall in a gravitational field.  $a^a$ is  the acceleration the observer has to exert to stay static by following a timeline killing 
field.
%\footnote{Note that we aim to deduce the inertial mass, so we don't define it initially.  We do assume that such a body has a Compton wavelength, $\lambda_C $ which provides the minimal scale within which it can be localized.}.

Note that while this TEP extends EP2, there is no extension of EP1 to this regime.  This is because the phenomena described by the TEP are on scales larger than $R$ and are thus not compatible with the restriction required by EP1 that the relevant phenomena be small compared to $R$.  Indeed, 
it appears
there is no simple way to characterize the spectrum that freely falling or orbital detectors observe in asymptotically deSitter spacetimes\footnote{I am thankful to conversations with Jurek Kowalski-Glikman on this and related questions.}.

\subsection*{Phase transition versus phase boundary}

The deSitter temperature is the minimum temperature that can be measured in equilibrium in a static spacetime with positive cosmological constant.  Similarly, the peak wavength is bounded above by the cosmological horizon,
\f
\lambda_{DL} \leq R.
\ff
 While we see many stars with
$a < a_\Lambda$, in terms of temperature and peak wavelength it appears that the cosmological constant dominated regime lives in the neighbourhood of a phase boundary characterized by approaching the minimal temperature or the maximal peak wavelength. 

This may then be a kind of quantum critical phenomena, at a minimal temperature, which is as close as an observer in deSitter spacetime can get to $T=0$.

If so we may expect phenomena in this region to scale.  Indeed, Milgrom proposes that $MOND$ behaviour is characterized by invariance under scaling\cite{MM-scale}.  He proposes that 
in the regime of small acceleration, 
equations of motion are to scale uniformly
under
\f
x^i \rightarrow \lambda x^i, \ \ \ \ \ \ t \rightarrow \lambda t
\label{scaling}
\ff
with masses held fixed, as are fixed constants including $\Lambda, G$ and $a_0$.  ($c$ is automatically invariant, while $\hbar$ requires a separate discussion, which we postpone as the factors of $\hbar$ cancel in the limit we are discussing here.  )  
%Energies and momenta scale as $\lambda^{-1}$, as do accelerations.

We note that Newton's law, 
\f
a_N^{\hat{r}} = - \frac{m_g}{m_I} \frac{GM}{r^2} 
\ff
is not homogeneous under scale transformations, so long as $m_g=m_I$.

We can use the possibility that $m_g = Z^{-1} m_I$ to modify the acceleration law to make it
scale invariant in the $\Lambda$  dominated regime.

Since $Z$ characterizes the $\Lambda$ dominated phase, we expect it may depend on $\Lambda$, but as it is dimensionless, it will have to depend on the ratio of $R$ to other
lengths.  A length of interest is $l_{a}$, hence we expect\footnote{Note here that $a$ is the magnitude of the acceleration of the static observer. }
\f
Z(\frac{R}{l_{acc}} ) = Z (\frac{T_U}{T_{dS}} ) = Z(\frac{a}{a_\Lambda} ) 
\ff
We note that when we express $Z$ in terms of accelerations, the $c$'s cancel.
If we impose scale invariance, then $Z$ must scale like $\lambda^{-1}$.  Hence scale 
invariance requires
\f
Z(\frac{a}{a_\Lambda} ) =\frac{a}{a_\Lambda} 
\ff
This gives us the MOND acceleration relation.
\f
a^i \frac{a}{a_\Lambda} = \nabla^i  \phi_N = - \hat{r}^i  \frac{GM}{r^2}
\ff

Note that the original observation which motivated this whole story, of flat rotation curves, is a verification of scaling, as the velocity, which is scale invariant, becomes independent of radius.

\section{Entropic definition of inertial mass}

How could there arise, from fundamental theory, a scaling of the ratio of inertial to gravitational mass?  
To explore this,  we bring in the entropic gravity hypothesis of 
Verlinde\cite{EVentropic}\footnote{The general relativistic version of this idea is described in \cite{Ted95,Ted2015,paddy}. A version of Verlinde's argument valid in loop  quantum gravity is in \cite{ls-entropy}.  Other proposals to derive $MOND$ from entropic gravity are
described in \cite{EVdark,vanP,AL}}.  However, we interpret Verlinde's idea as an entropic elucidation of the concept of-and principle of-inertia.  The idea is that the inertial mass arises from consideration of how well a particle can be localized.

Following \cite{EVentropic,vanP}, we assign an entropy to how localizable a particle is by a static detector confined within an horizon.   The minimal localization a static, accelerating detector can make of a particle is that it is within some distance scale, $L$, defined by a detector.  The detector should be smaller than her horizon, so $L \leq R$.  The best localization of the 
particle she can make is that the particle is within a Compton wavelength, $\lambda_C$.  We can then define an entropy of localization which is the negative of  the information gained by such a localization, which is proportional to the ratio of the best and worst possible localization,
\f
S_{loc}= 2 \pi \frac{L}{\lambda_C}
\label{Sloc}
\ff
This entropy counts the information potentially available by localizing the particle.  

First we check that this gives Newtonian dynamics when the cosmological constant is turned off, which means we take $R \rightarrow \infty$ and $T_{DL} = T_U$.
The corresponding free  energy is (for vanishing cosmological constant).
\f
W= T S = \frac{\hbar a}{ 2 \pi c} 
S_{loc}= \frac{L}{\lambda_C} \frac{\hbar a}{c}
\ff

Now we follow Verlinde in defining an entropic force as $T$ times the differential of $S$ gotten by moving the particle a distance $\Delta L$ within by the detector,
\f
F= T\frac{\Delta S}{\Delta L} = \frac{\hbar  }{\lambda_C c} a 
\ff
We can now {\it define} the inertial mass in terms of the Compton wavelength, by
\f
m_I \equiv \frac{\hbar }{\lambda_C c}
\ff
to find Newton's second law.
\f
F= m_I a
\ff
Hence, we have derived the principle of inertia, and inertial mass, from entropic considerations.

%We can also derive the rest energy as a thermodynamic energy, by integrating the force over distance
%\f
%U_{rest}= \int_0^{L_T}  dx F = m_I a L_T = m_I c^2
%\ff
%where we make use of the fact that the peak wavelength is related to the acceleration by
%\f
%L_T = l_a= \frac{c^2}{a}
%\ff
%But the rest energy is what appears in the energy momentum tensor so it is the passive 
%gravitational mass, $m_g$.  
%\f
%U_{rest}= m_g c^2 = m_I c^2
%\ff
%So we have established the equality of gravitational and inertial mass.  

But this argument assumed the cosmological constant vanishes.  What happens when we turn a small
$\Lambda$ on?  We consider the case of positive $\Lambda$, in which case there is a cosmological horizon,  at a distance $R$ defined by $\frac{\Lambda}{3}= \frac{1}{R^2}$. 

In the presence of $\Lambda$ there is an irreducible deSitter temperature
$T_{dS}$.  Because the temperature in equilibrium cannot be reduced to below, $T_{dS}$,
hence we can posit that the change in free energy is,
\f
\Delta W = \left ( T_{DL}- T_{dS} \right ) \frac{\Delta S}{\Delta L}  
\ff
It is necessary to subtract off $T_{dS}$ in the entropic derivation of the force law, otherwise
a freely falling detector with $a=0$ would experience a force.

To get the force we take again the derivative with respect to $\Delta L$ to find
\f
F=  \left ( T_{DL}- T_{dS} \right ) \frac{\Delta S}{\Delta L} = 
\frac{a^2}{2 a_\Lambda }
\frac{\hbar  }{\lambda_C c} 
\ff
%This gives
%\f
%F= m_I a \left ( \frac{a}{2 a_\Lambda} \right )= m_I a Z ( \frac{a}{2 a_\Lambda}  )
%\ff
This should be set equal to the Newtonian force law, $F=m_g a_N$

Assuming that $m_g=m_I$, this leads to the MOND dynamics
\f
%m_I a = m_g a_N, \ \ \ \ \rightarrow 
a^2 = 2 a_N a_\Lambda
\ff

%Now suppose that $T$ is so small that the peak wavelength doesn't fit into the cosmological horizon. Then, in the computation of the rest energy,we find
%\f
%U_{rest, \Lambda}= \int_0^{l_a}  dx F = m_I a R = m_I c^2 Z
%\ff
%So we have
%\f
%m_g= m_I c^2 Z=  m_I c^2 \frac{T_\Lambda}{T}=  m_I c^2 \frac{a_\Lambda}{a}
%\ff
We can explain this by noting that when there is no cosmological horizon the entropic force is proportional
to $T_U  \propto a$, whereas in the presence of a positive cosmological constant we have, for small $a$,
\f
F \propto (T_{DL}- T_{dS} )  \propto  \frac{a^2}{2 a_\Lambda}
\ff

Alternatively,  the deviation from Newtonian physics seen in MOND can be expressed by a renormalization of the relation between inertial and gravitational mass, inserted into the standard Newtonian laws of gravity and inertia.
\f
m_I = m_g Z [ \frac{a}{a_\Lambda}],
\ff
with $Z [ \frac{a}{a_\Lambda}] = \frac{a}{a_\Lambda}$ for $a < a_\Lambda$. 

Notice several things:  The temperature is always that seen by a static observer.  This is an accelerated observer, locally.  It is not on a geodesic, so the acceleration is non-zero for Newton, special and general relativity.  The static observer may observe particles in free fall and in orbit.  We have no reason to worry about what temperatures those particles would see were they equipped with detectors.  This explains how the results can be valid for circular motion as seen in spiral galaxies.

Second, the thermodynamics we employ is equilibrium thermodynamics.  There must be periods of non-equilibrium behaviour, while a galaxy is forming.  Our simple model has so far nothing to say about such behaviour, it assumes the galaxy has been static for a long time.

Third, from this argument we recover the scaling (\ref{scaling}) in the cosmological constant dominated regime.   This supports the picture that $MOND$ is a kind of critical phenomena related to being near the boundary of minimal temperature  at $T= T_{dS}$.

Finally, this entropic  origin of inertia is reminiscent of the idea that inertial motions are those that see the least thermal fluctuations\cite{fluctuations,KP}.

\section{Effective description}

The conclusion we come to, from the forgoing, is that $MOND$ can be described as a 
modification of the principle of inertia, arising from the insertion, into the ratio of gravitational to inertial mass, of a renormalization
factor, $Z$, which is a function of global or environmental variables.
So the equations of motion for a star in a galaxy, with trajectory $x^i (t)$ becomes modified to 
\f
\ddot{x}^i = Z^{-1} g^{ij} \nabla_j \phi
\label{A1}
\ff
where $g^{ij}$ is the inverse of the spatial metric. $Z$ can be read as
\f
Z= \frac{m_I}{m_g} = Z (\frac{T_U}{T_{dS}}) = Z(\frac{a}{a_\Lambda } )
\ff
where $T_U$ is the Unruh temperature 
observed by a static detector held at that point, and $a$ is the magnitude of the acceleration necessary to hold that detector in place.

We can express this as a bi-metric theory, in which the metric that governs the motion of particles is not the same as the metric that solves the Einstein equations.

We note that for a particle in free fall or orbital motion  in the static gravitational field, $\ddot{x}^i = - a^i$, by the equivalence principle (where $a^i$ is evaluated at their location).  But these are conceptually distinct and generally not equal.  $a^i$ is the acceleration of a static detector held at a fixed position in the gravitational field, while $\ddot{x}^i$ is the acceleration of a particle in free fall.  Moreover, the dependence on $a^i$ is actually a proxy for a dependence of the ratio of gravitational to inertial mass on the local temperature.  Thus, we do not have an issue of equations of motion of higher order in time derivatives.  We do have a question as to how the dependence of $T_U$ evolves in time in non-static configurations or non-equilibrium states\footnote{As a result of the limitation to motion in static spacetimes,  we are unable to address the issue of instabilities.}.

In earlier sections, I argued that both scale invariance and a thermodynamic or entropic origin of the first law suggests
\f
 Z (\frac{a}{a_\Lambda } )= \frac{a}{a_\Lambda } \ \ \ \mbox{for } \ \ \ a < a_\Lambda
\ff
and $Z=1$ otherwise.

Let's next see if we can describe this phenomena in terms of an effective picture in the language of general relativity.  We can, in a limited sense that applies only to static spacetimes, as follows.

We start with (\ref{A1}), which tells us that we can describe $MOND$ as a replacement,in the spherically symmetric case
\f
g^{rr} \partial_r \phi \rightarrow Z^{-1} g^{rr} \partial_r \phi
\ff
Recall that in the Newtonian limit of general relativity
\f
g_{00} = 1-\frac{2 \phi}{c^2} =f
\ff
This suggests that the metric that governs the motion of particles in a static gravitational field is
modified by
\f
g_{rr} \rightarrow \tilde{g}_{rr}= Z g_{rr}, 
%\ \ \ \ \ \tilde{g}_{00}= g_{00}, \ \ \ \ \ \tilde{g}_{0i}= g_{0i}
\ff
with the rest unmodified.

We may call this the {\it thermal metric.}\footnote{This name was suggested by Matteo Smerlak.}.
If we recall that the leading term in the geodesic equation, in the Newtonian limit of the spherically symmetric case, is
\f
\ddot{x}^r = \Gamma^r_{00} \dot{x}^0 \dot{x}^0 \approx \Gamma^r_{00}c^2
\ff
where
\f
\Gamma^r_{00}=- \frac{1}{2}  g^{rr} \partial_r g_{00}
\ff
This gives to leading order
\f
\Gamma^r_{00} \rightarrow \tilde{\Gamma}^r_{00} = Z^{-1} \Gamma^r_{00}
\ff

Then we have a prescription for encoding $MOND$ as an effective description within general relativity, in the case that the metric is static.  
\begin{enumerate}

\item{} Leave the field  equations in the original metric, $g_{ab}$ unmodified.

\item{}The motion of a particle in a static gravitational field is given by the geodesics of the modified, thermal metric, i.e. from the variational principle
\f
S =m_I \int ds \sqrt{\tilde{g}_{ab} \dot{x}^a \dot{x}^b }
\label{Sp}
\ff
where the metric  $\tilde{g}_{ab}$ is defined by the radial component being rescaled
by $g_{rr} \rightarrow \tilde{g}_{rr}= Z[ {\cal F} ] g_{rr}$.

\item{} We then study the variational principle,
\f
\delta_{\cal F} S =0
\label{var}
\ff
The meaning of $\delta_{\cal F}$ is that
when varying the action, the path is to be varied with ${\cal F}$ held as a fixed function.  This gives us $x^a (s)_{\cal F}$, where there is a solution for every fixed function, ${\cal F}$. We are then interested in those solutions that satisfy\footnote{In the MOND limit.},  (\ref{var}) together with
\f
{\cal F}= \frac{a}{a_\Lambda }
\label{consistency}
\ff
This may be thought of as a condition of  equilibrium.  The idea is that $Z$ is defined as  a function of ratios of temperatures, as a result of processes of equilibration to the Deser-Levin temperature.  In the MOND, or cosmological constant dominated regime, this is close to the deSitter temperature, so equilibration take place over cosmological time scales.  The particle motion takes place over much shorter time scales, and is determined by the variational principle with respect to which the thermal background, and hence $Z$, can be considered fixed.  (\ref{consistency}) is a consistency condition that expresses the fact that the dynamical and thermal equilibrium of the  galaxy is made up of vast numbers of stars.

\end{enumerate}

This reproduces, to zeroth order in $\frac{v^2}{c^2}$ the picture we arrived at above.  There is, however, an additional term in the modified geodesic equation which survives the non-relativistic limit $c \rightarrow \infty$. It  comes from the term in the geodesic equation
$\tilde{\Gamma}^r_{rr} \dot{r}^2$.  Including it,  the full radial part of the geodesic equation reads,
\f
\ddot{x}^r =- \frac{1}{Z} g^{rr} \partial_r \phi - \frac{\dot{r}^2}{c^2} \partial_r \phi 
- \frac{\dot{r}^2}{\partial_r \phi }2 \pi G \rho \ \Theta [ Z-1   ]
\label{ddx}
\ff
%where $r_0 =  \sqrt{ R R_{Schw}}$ is the critical radius outside of which MOND turns on.
We also use the fact that
\f
Z =\frac{| \partial_i \phi |}{a_0}
\ff
where $\nabla^2 \phi = 4 \pi G\rho$.  
The second term is standard from the Schwarzschild solution.  The third term is novel.  It is small for most galaxies because, because it vanishes for circular motion. In addition, by the time the theta function turn on, indicating we are in the $MOND$ regime, the baryon density will be falling off exponentially.  It will make a contribution to the precision of the perihelion for the orbits of stars in disk galaxies, but these do not seem to be easy to measure.   But it still may be the source of new effects that might be observable.  More worryingly, it may have a non-trivial effect on the radial motions of stars in elliptical galaxies.

\subsection{Hamiltonian analysis}

To understand better the modification in dynamics implied by (\ref{ddx}) we turn to a Hamiltonian
analysis of particle motion in the thermal geometry. 
We proceed as usual\cite{Wald}  and identify two conserved quantities for a particle with trajectory $x^a (s)$ and
four velocity $u^a = \frac{dx^a}{ds}$,
\f
E=\tilde{g}_ab u^a k^b = f \dot{t}, \ \ \ \ \ L=\tilde{g}_ab u^a r^b =r^2 \dot{\phi}
\ff
where $k^a$ and $r^a$ are the time and angular Killing vector fields.
Setting $-1 = \tilde{g}_ab u^a u^b$ yields the conservation law,
\f
E^2 = \frac{Z}{2} \dot{r}^2 +V
\ff
where $V$ is the standard Schwarzschild potential,
\f
V =\frac{1}{2}  \left [  
1 - \frac{2 G M}{r} + \frac{L^2}{r^2} - \frac{2 GML^2}{r^3}
\right ]
\ff
The variational principle (\ref{Sp}) is then equivalent to a lagrangian
\f
{\cal L}= \frac{Z}{2} \dot{r}^2 -V
\ff
We turn this into a Hamiltonian in the usual way,
\f
{\cal H}= \frac{p^2 }{2Z}+ V,
\ff
where the radial momentum is defined by
\f
p = \frac{\delta {\cal L}}{\delta \dot{r}} = Z \dot{r}
\ff
We also have
\f
\dot{p} =- \frac{\delta {\cal H}}{\delta r}= - \frac{\partial V}{\partial r} + \frac{p^2}{Zr} \theta [1-Z]
\ff
from which we deduce the radial acceleration relation,
\f
\ddot{r} = - \frac{1}{Z}\frac{\partial V}{\partial r} - \dot{r}^2 \frac{2 \pi G \rho}{\partial_r \phi} \theta [1-Z]
\ff
where we have used Using $Z=\frac{\partial_r \phi }{a_0}$.  This reproduces (\ref{ddx}).

Outside the matter distribution, where $\phi = \frac{GM}{r}$, this can also be expressed as
\f
\ddot{r} = - \frac{1}{Z}\frac{\partial V}{\partial r} - \frac{\dot{r}^2}{r}  \theta [1-Z]
\ff

\section{Discussion}

We have arrived, very tentatively, at a kind of  effective, bi-metric description, so far valid only for static spacetimes.  The  standard metric, $g_{ab}$ is a solution to Einstein's equations, whose source is baryonic matter.  However, the metric that governs the motion of particles is
a different metric, $\tilde{g}_{ab}$, which differs from the standard metric only by a scaling of the radial component by a function of the temperature seen by a static observer.

This is justified by a rough argument based on the  entropic gravity hypothesis of Verlinde and others, modified by the presence of a positive cosmological constant.  It can also be justified by a scaling argument based on the scaling hypothesis of Milgrom.  That hypothesis can in turn perhaps be justified by seeing the MOND regime as critical phenomena due to the proximity 
of a phase boundary.

It is painfully clear that the ideas discussed here are tentative and unlikely.  Surely dark matter is a far simpler and less challenging hypothesis, and we will all breath a sigh of relief if and when it is detected.  Meanwhile, there are several challenges and oportunities which are raised by the present proposal.

\begin{itemize}

\item{}The  far field is a big issue.  $MOND$ can't go on forever, for one thing we need to make contact with the large scale successes of general relativity.  $Z$ needs to level off at a constant at some distance, 
hopefully with a scaled mass. Until this is addressed the theory has nothing to say about  the relevance of MOND scaling at extragalactic scales relevant for clusters and large scale structure, where dark matter hypothesis appears to work well.

\item{}Quantum gravity must ultimately explain the effective bi-metric structure.  One possibility is to employ disordered locality\cite{disordered,ChandaLee} as studied in\cite{Leeun}.

\item{}In the above analysis we assumed disk galaxies are spherical, which they are not.

\item{}We assumed that there is an immediate transition between $Z(\frac{a}{a_0} )$ between
$z=1$ for $a > a_0$ and $Z= \frac{a}{a_0} $ when $a < a_0$.  In most of the MOND literature this transition is softened by an interpolating function as in (\ref{function}).

\item{}Most interestingly, we have found a new prediction for dynamics in the $MOND$ regime,
given by the third term in (\ref{ddx}).  This does not affect circular motion, but it does introduce a damping of radial motion which may play a role in the dynamics of elliptical galaxies or in the formation of disk galaxies.  We note that it is present only in the MOND regime, but it does not depend on the value of $a_0$.  

\item{}What we have studied here is an effective theory, limited to test matter propagating on static spacetimes.  The next step is to stationary spacetimes and should be straighforward.
Beyond that we need to consider how to consistently couple matter to the gravitational field
in the MOND regime. Equivalently, we want to consistently couple the two metrics to each other and to matter.   If the picture presented here is in the right direction,  that dynamics will be non-local and out of  equilibrium.  It is unlikely this can be expressed in a local or field theoretic form.  Indeed, it is possible that the processes of equilibration on a cosmological scale reveal irreversible aspects of quantum gravity,
as discussed in \cite{SURT} - \cite{TA2}.

\end{itemize}

\section*{ACKNOWLEDGEMENTS}

I am especially grateful to Mordehai Milgrom, Stacy McGaugh,  Jurek Kowalski-Glikman, Sabine Hossenfelder, Joao Magueijo,  Maurice van Putten and and Yigit Yargic for advice, encouragement, comments on drafts and correspondence.  I would like also to thank Andrzej Banburski, Jacob Barnett,  Joseph Bramante,  Lin-qing Chen,  Marina Cortes, Bianca Dittrich, Laurent Freidel,  Henriques Gomes,   Andrew Liddle,  John Moffat,  Krishnamohan Parattu, Percy Paul, Vasudev Shyam and Matteo Smerlak  for very helpful discussions.   I am also
indebted to Stacy McGaugh for permission to reproduce Figure 1 and its caption from \cite{MLS}.  I also thank a referee for helpful suggestions.  I would especially like to thank Stephon Alexander for collaboration in ongoing work, which will be reported elsewhere.

This research was supported in part by Perimeter Institute for Theoretical Physics. Research at Perimeter Institute is supported by the Government of Canada through Industry Canada and by the Province of Ontario through the Ministry of Research and Innovation. This research was also partly supported by NSERC and FQXi and by a generous grant from the John Templeton Foundation.

\end{document}